\documentclass[%
 reprint,
superscriptaddress,
 amsmath,amssymb,
 aps,
]{revtex4-2}

\usepackage{graphicx}
\usepackage{dcolumn}
\usepackage{bm}

\begin{document}

\preprint{APS/123-QED}

\title{{Electrically-driven Photonic Crystal Lasers with Ultra-low Threshold}}

\author{Evangelos Dimopoulos}
\affiliation{DTU Electro, Department of Electrical and Photonics Engineering, Technical University of Denmark, Ørsteds Plads 343, Kgs. Lyngby, DK-2800,
Denmark.}

\affiliation{NanoPhoton - Center for Nanophotonics, Technical University of Denmark, Ørsteds Plads 343, Kgs. Lyngby, DK-2800,
Denmark.}

\author{Aurimas Sakanas}
\affiliation{DTU Electro, Department of Electrical and Photonics Engineering, Technical University of Denmark, Ørsteds Plads 343, Kgs. Lyngby, DK-2800,
Denmark.}

\author{Andrey Marchevsky}
\affiliation{DTU Electro, Department of Electrical and Photonics Engineering, Technical University of Denmark, Ørsteds Plads 343, Kgs. Lyngby, DK-2800,
Denmark.}

\author{Meng Xiong}
\affiliation{DTU Electro, Department of Electrical and Photonics Engineering, Technical University of Denmark, Ørsteds Plads 343, Kgs. Lyngby, DK-2800,
Denmark.}
\affiliation{NanoPhoton - Center for Nanophotonics, Technical University of Denmark, Ørsteds Plads 343, Kgs. Lyngby, DK-2800,
Denmark.}

\author{Yi Yu}
\affiliation{DTU Electro, Department of Electrical and Photonics Engineering, Technical University of Denmark, Ørsteds Plads 343, Kgs. Lyngby, DK-2800,
Denmark.}
\affiliation{NanoPhoton - Center for Nanophotonics, Technical University of Denmark, Ørsteds Plads 343, Kgs. Lyngby, DK-2800,
Denmark.}

\author{Elizaveta Semenova}
\affiliation{DTU Electro, Department of Electrical and Photonics Engineering, Technical University of Denmark, Ørsteds Plads 343, Kgs. Lyngby, DK-2800,
Denmark.}
\affiliation{NanoPhoton - Center for Nanophotonics, Technical University of Denmark, Ørsteds Plads 343, Kgs. Lyngby, DK-2800,
Denmark.}

\author{Jesper Mørk}%

\affiliation{DTU Electro, Department of Electrical and Photonics Engineering, Technical University of Denmark, Ørsteds Plads 343, Kgs. Lyngby, DK-2800,
Denmark.}
\affiliation{NanoPhoton - Center for Nanophotonics, Technical University of Denmark, Ørsteds Plads 343, Kgs. Lyngby, DK-2800,
Denmark.}

 \author{Kresten Yvind}%
 \email{kryv@dtu.dk}
\affiliation{DTU Electro, Department of Electrical and Photonics Engineering, Technical University of Denmark, Ørsteds Plads 343, Kgs. Lyngby, DK-2800,
Denmark.}
\affiliation{NanoPhoton - Center for Nanophotonics, Technical University of Denmark, Ørsteds Plads 343, Kgs. Lyngby, DK-2800,
Denmark.}

\date{\today}

\begin{abstract}

Light sources with ultra-low energy consumption and high performance are required to realize optical interconnects for on-chip communication. Photonic crystal (PhC) nanocavity lasers are one of the most promising candidates to fill this role. In this work, we demonstrate an electrically-driven PhC nanolaser with an ultra-low threshold current of 10.2 \textmu A emitting at 1540 nm and operated at room temperature. The lasers are InP-based bonded on Si and comprise a buried heterostructure active region and lateral p-i-n junction. The static characteristics and the thermal properties of the lasers have been characterized. The effect of disorder and p-doping absorption on the Q-factor of passive cavities was studied. We also investigate the leakage current due to the lateral p-i-n geometry by comparing the optical and electrical pumping schemes.

\end{abstract}

\maketitle


\section{Introduction}
Information and Communication Technologies (ICT) are flourishing, delivering life-changing services that necessitate a rising amount of data processing, storage, and communication. In the past decade,
ICT services have grown at an exponential rate \cite{IEA21_exponentialDataCenters}, and are expected to account for a significant portion of global electricity consumption by 2030 \cite{Andrae2015}, necessitating novel energy-efficient solutions. Transitioning from electrical interconnects to optical interconnects has enabled much improvement since optical communication has advantages in terms of bandwidth, speed, and power consumption and has been already employed for short-distance communication in data centers and supercomputers via vertical-cavity surface-emitting lasers (VCSELs). Extending this concept for chip-to-chip and on-chip communication will be a game-changer to ICT, however, conventional light sources cannot meet the low power consumption requirement \cite{Miller1997}.

Photonic crystal (PhC) nanolasers have been developed by various groups and have showcased great potential and rich physics \cite{Matsuo2010,Hamel2015,Yu2017,Yu2021}, however, most these studies were limited to optical pumping, while electrical operation was achieved under pulsed pumping \cite{Park2004}, or cryogenic temperatures \cite{Ellis2011_Electrical_QDs_cryogenic}. Recently, this steep technological barrier has been overcome and monolithic,\cite{Seo2007, Takeda2013, Jeong2013_electrical_NanoBeam} and heterogeneously-integrated \cite{Takeda2014_heterogeneously_integrated, Crosnier2017, Takeda2021} electrically-pumped PhC lasers achieved continuous-wave room-temperature operation. However, there is significant room for improvement in terms of power consumption and energy efficiency and thus further research is essential.

In this work, we investigate the properties of electrically-injected PhC nanolasers with a wavelength-scale active region that are operated in continuous-wave at room temperature and are heterogeneously integrated on Si via direct bonding. 2D PhC membranes are a favorable platform for transferring to non-native substrates (such as Silicon), for high-resolution lithography, and local doping, leading to an integrated structure that allows current injection through a lateral junction. All these properties are critical for CMOS compatibility enabling hybrid photonic-electronic integration. In this paper, we demonstrate a PhC nanolaser with an ultra-low threshold current of 10.2 \textmu A emitting at 1540 nm, and we characterize the thermal properties of PhC lasers with one and three quantum wells (QWs). Furthermore, the quality factor (Q-factor) of passive PhC cavities was investigated, and the optical losses due to p-doping were determined. Finally, the problem of injection efficiency drop is examined by comparing the behavior of a laser under optical and electrical pumping.

\section{Design and Fabrication}

\subsection{Device Structure}
\begin{figure*}
  \centering\includegraphics[width=0.8\textwidth]{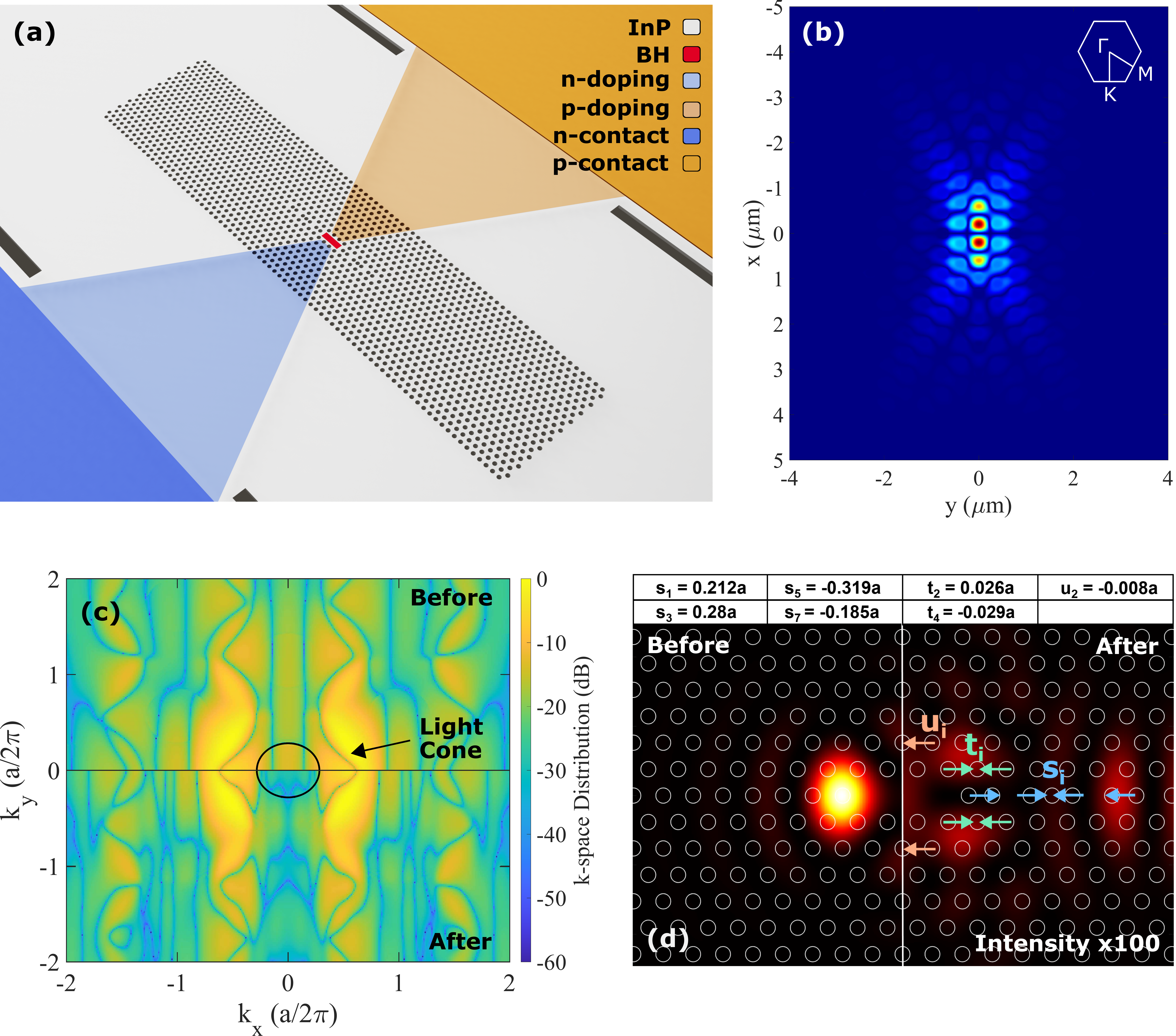}
\caption{(a) Schematic diagram of the laser. (b) The magnetic field distribution of the fundamental mode of an optimized L3 PhC cavity. (c) Two-dimensional Fourier transform of the $E_y$ in reciprocal space plotted in logarithmic scale before (top) and after (bottom) the Q-factor optimization. The solid black circle indicates the light cone. (d) Visualization of the leaky components of cavity before (left) and after (right) the Q-factor optimization. The hole position tuning is denoted as $s_i$, $t_i$ and $u_i$ for the center, first and second horizontal PhC row respectively.}  \label{fig1_design}
\end{figure*}

A schematic of a Line Defect (LD) PhC laser is given in \textbf{Figure \ref{fig1_design}}a. The LD cavity is used to confine photons and is created by omitting some holes along the $\Gamma-K$ direction of a triangular PhC lattice. The active medium consists of a buried heterostructure (BH) containing one or three InGaAsP/InAlGaAs
 QWs, and it is placed inside the PhC cavity confining the carriers. The carriers are injected into the BH region by a lateral \textit{p-i-n} junction scheme.

To attain a lasing wavelength in the C-band while maintaining a high Q-factor, the PhC lattice constant, the radius, and the InP slab thickness were chosen as 440 nm, 120 nm, and 250 nm respectively (see also Figure S1, Supporting Information). The hole radius is varied across devices to compensate for variations of the membrane thickness across the wafer and provide a better overlap between the cavity resonance and the photoluminescence peak of the active medium. In this work, the cavity designs are standard (holes removed from uniform lattice) LD cavities \cite{Saldutti2021} of different lengths. The L3 cavities were modified by adjusting the position of the surrounding holes to increase the Q-factor and achieve lasing.

In Figure \ref{fig1_design}b, the magnetic field of the fundamental mode of an L3-optimized cavity is calculated via the three-dimensional finite-difference time-domain (FDTD) method. To achieve a high Q-factor, the spatial frequency components that lie within the cone of light need to be minimal \cite{Akahane2003}. In Figure \ref{fig1_design}c the two-dimensional Fourier transformation of the $E_y$ field is shown before and after optimization. The Q-factor was optimized by visualizing the leaky field components in real space using a subsequent inverse Fourier transformation and adjusting the position of the holes around the leaky area \cite{Nakamura2016}. The leaky field profile before and after optimization can be seen in Figure \ref{fig1_design}d. Because the leaky field was suppressed after optimization, its intensity had to be scaled up by a factor of 100 to be comparable to the non-optimized one. The position shift of the i\textsubscript{th} hole of the center, 1\textsubscript{st} and 2\textsubscript{nd} PhC row is denoted as $s_i$, $t_i$ and $u_i$ respectively and their values are included in the figure. Using this intuitive optimization method the Q-factor was improved from 5000 to $1.1\cdot 10^6$ by tuning the position of seven PhC holes.

\begin{figure*}
\centering\includegraphics[width=0.9\textwidth]{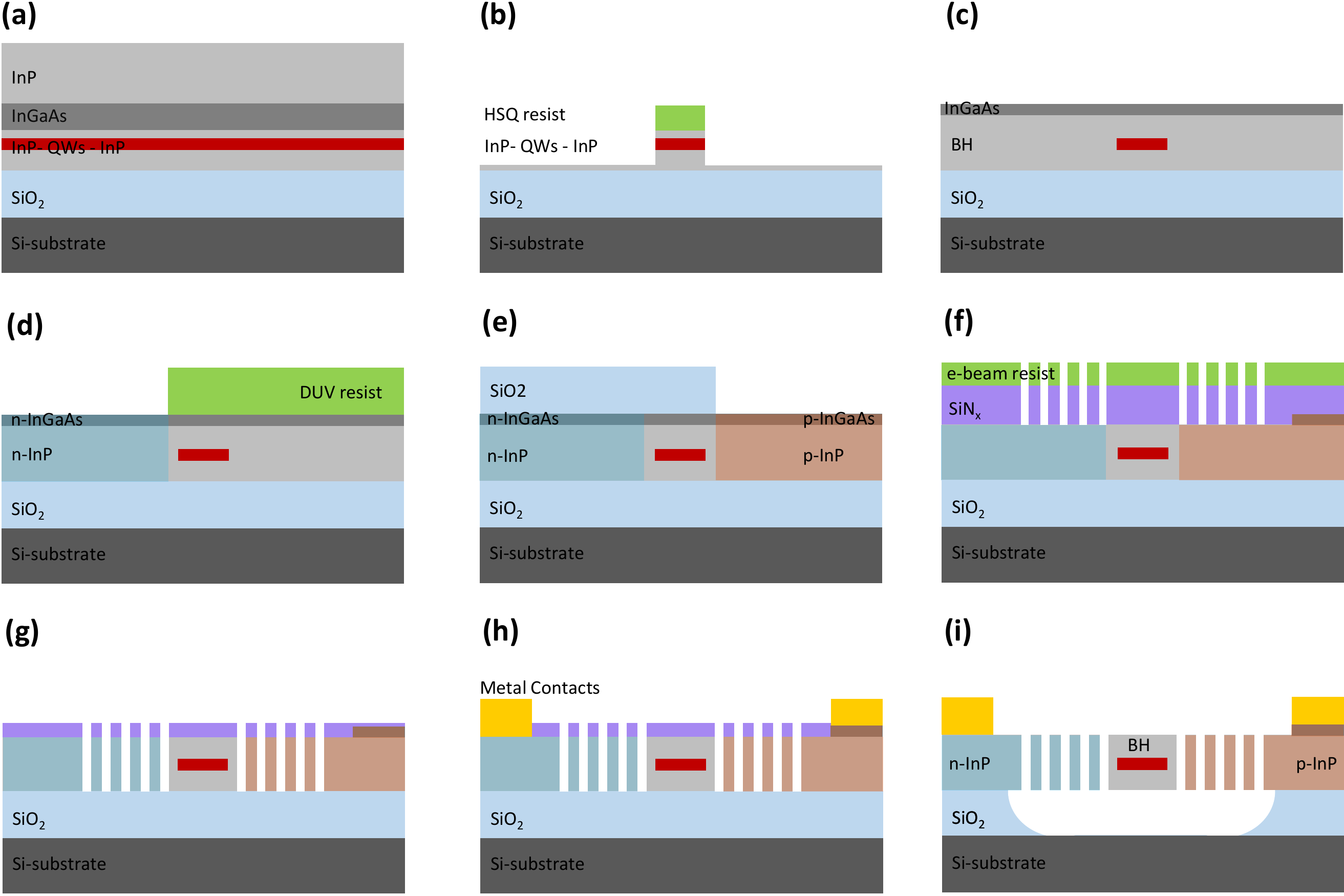}
\caption{Fabrication process of the device. (a) Directly-bonded InP wafer on Si/SiO2 wafer. (b) HSQ mask-protected mesa-structures formed after the dry wafer etching. (c) Buried-Heterostructure formation after the second regrowth of InP. (d) Wafer after Si-ion implantation before removing the DUV mask used for n-doping. (e) p-doping using Zn thermal diffusion. (f) E-beam lithography for the PhC holes definition. (g) Wafer after the two-step dry etching.  (h) Metallizations using lift-off processes. (i) Membranization of PhC structures. }  \label{fig2_fabrication}
\end{figure*}

\subsection{Fabrication Process}
The following part describes the laser device fabrication, provides explanations for some of the processing choices, and highlights the main effects that fabrication imperfections can have on the resulting laser performance. Some of the process steps are depicted in Figure \textbf{\ref{fig2_fabrication}}.

Precise alignment between the buried heterostructure, photonic crystal cavity, and doped p- and n- regions is the essential requirement for the device operation. While the BH and the PhC holes are defined by electron beam lithography, we chose deep ultraviolet (DUV) lithography to define doping regions because of their overall much larger dimensions. Since our DUV tool imposes a limitation on the minimum diameter of the wafer, we use a 4” silicon with a thermal oxide layer as a carrier wafer onto which we directly bond 2" InP epiwafer. The use of even larger diameter wafers is mostly restricted by the unavailability of suitable III-V processing equipment in our cleanroom facility.

Laser device processing begins with the following III-V-on-Si integration procedure: Epitaxially grown etch-stop InGaAs and InGaAsP/InAlGaAs QW layers on the 2” InP wafer are directly bonded to the middle part of the 4” Si wafer with 1100 nm thermal oxide [Figure \ref{fig2_fabrication}a]. The direct bonding is facilitated by a thin intermediate Al\textsubscript{2}O\textsubscript{3} layer for improved bonding strength \cite{Sahoo2018}.
Then, the InP substrate is removed by chemical etching in HCl terminating at the etch-stop layer, which is removed by another chemical etching in H$_2$SO$_4$:H$_2$O$_2$:H$_2$O mixture.

Before the actual device fabrication, we form the alignment marks in Si. First, a combination of optical lithography together with dry and wet etching is used to selectively remove InP and SiO$_2$ in dedicated wafer areas. Then, the marks for the e-beam and DUV alignment are exposed by optical contact lithography and etched into Si with SF$_6$/O$_2$ chemistry. The alignment precision between the features formed by the e-beam and DUV is thus fundamentally limited by the laser writer precision with which the physical mask was manufactured. For multiple e-beam exposures, our estimates indicate that the wafer level statistical 3-sigma standard deviation is below 50 nm and limited by the sidewall roughness of the alignment marks after dry etching, as well as wafer stress non-uniformity \cite{Sakanas2019}.

The alignment marks defined in the Si substrate are then used for aligning and exposing the buried heterostructure mask pattern into a high-resolution negative-tone hydrogen silsesquioxane (HSQ) resist by e-beam lithography. After development, the HSQ mask is used as a hard-mask in the following dry etching step with HBr/CH$_4$/Ar chemistry at elevated 180 $^{\circ}$C temperatures in an inductively-coupled plasma etcher. InP together with the QW layers is removed outside the mask-protected regions [Figure \ref{fig2_fabrication}b]. These etched InP/QW regions are then refilled by InP in an epitaxial selective-area MOVPE regrowth step. After this first regrowth, the HSQ mask is removed by HF-etching, and the second regrowth is used for the surface planarization and for defining the final III-V device layer thickness of 250 nm [Figure \ref{fig2_fabrication}c].

Electrical injection of carriers is realized via a lateral p-i-n doping configuration. Such configuration matches well the planar structure of the membrane laser without the need to increase its thickness by the regrowth of additional doped layers, which could compromise the crystal quality of the III-V-on-Si by exceeding the critical thickness \cite{Fujii2015}. On the other hand, achieving high carrier injection efficiency requires high-precision alignment of the doped regions to the BH. Aligning to the Si marks in DUV stepper, we first expose the set of openings to form the n-type doping regions via Si ion implantation [Figure \ref{fig2_fabrication}d]. The procedure is repeated to define openings for Zn diffusion to form p-type doping regions [Figure \ref{fig2_fabrication}e]. Technical details about the doping process are provided elsewhere \cite{Marchevsky2019}.

After the doping is complete, the mask design of the photonic crystal cavities is aligned and exposed in e-beam. The patterns are transferred in a two-step dry etching process from the e-beam resist to the SiN hard-mask deposited before the e-beam step [Figure \ref{fig2_fabrication}f], and then from the SiN hard-mask to the InP layer [Figure \ref{fig2_fabrication}g]. Any misalignment in this step between the BH and the PhC cavity would reduce the spatial overlap between the optical mode and the gain region, and in the extreme case, the holes would etch through the QW layers exposing them to air, which would inevitably result in significant non-radiative surface recombination \cite{Xue2015}.

Finally, the device fabrication is completed when the metal pads are formed on the n- and p-doped regions [Figure \ref{fig2_fabrication}h], and the PhC cavities are membranized by selectively HF-etching underlying thermal SiO2 layer [Figure \ref{fig2_fabrication}i].

\section{Results}
\subsection{Laser Characteristics}

\begin{figure*}
\centering\includegraphics[width=0.85\textwidth]{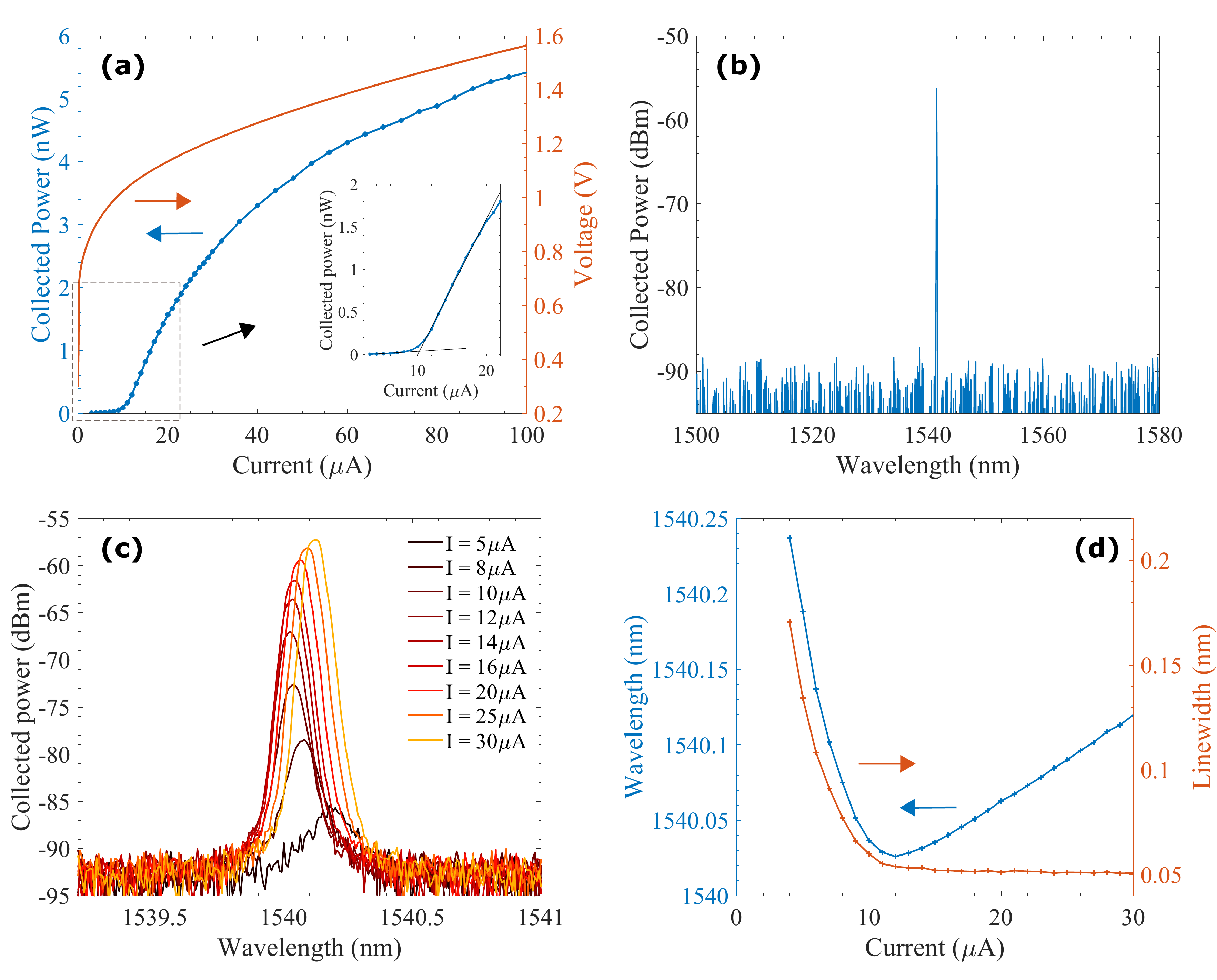}
\caption{Electrically-driven L3 PhC laser. (a) Collected output power and voltage versus injection current. The inset shows a close-up of the L-I curve at the region around the threshold. (b) OSA trace at 100\textmu A injection current. (c) The spectral evolution of the laser for different injection currents. (d) The peak wavelength and the linewidth of the emission peak as a function of the injection current.}\label{fig.L3}
\end{figure*}

The static properties of the PhC lasers were characterized at room temperature. The vertically scattered light from the cavity is collected using a 50x long-working distance objective, then it is coupled to a multi-mode fiber and is measured using an Optical Spectrum Analyzer (OSA). The Light-Current and the Current-Voltage curves of an optimized L3 cavity laser are shown in \textbf{Figure \ref{fig.L3}}a. The inset of \ref{fig.L3}a shows the output power around the characteristic transition from the spontaneous emission to stimulated emission, along with a two-line segmented fit used to calculate the threshold current of a laser. The device exhibits an ultra-low threshold current of 10.2 \textmu A.

The laser is single-moded, emitting at 1541 nm. In figures \ref{fig.L3}(b-c), the spectrum for different pumping currents is shown. In figure \ref{fig.L3}d, the spectral evolution of the laser is shown. In the spontaneous emission regime, the emitted wavelength blueshifts due to the carrier-filling effect \cite{Bennett1990_carrier_effect}. At and above the threshold, the quasi-Fermi levels are pinned and the wavelength redshifts due to heating induced by the high optical power density. Similarly, the linewidth of the laser saturates at the threshold, reaching the resolution limit of the OSA.

In this first generation of devices, an output waveguide was omitted to increase the device density, and thus, the total number of design variations, in order to study and fine-tune the different parameters affecting the laser performance. This also eliminates the need of decoupling the intrinsic Q-factor from a loaded Q-factor, however, a direct measurement of the slope efficiency and wall-plug efficiency was not possible, since it is difficult to estimate the actual output power. Evanescent coupling to a Si output waveguide will be included in future designs.

\subsection{The effect of disorder and p-doping} \label{ch.Q-factor}
An important parameter of a laser is the Q-factor of the laser cavity, which quantifies the temporal confinement of the photons. PhC cavities with ultra-high Q-factors exceeding one million have been demonstrated on the Si platform \cite{Asano2017_NTT_11M_Qfactor, Lai2014_Galli_2M_L3_Qfactor}. For InP-based PhC cavities, however, fabrication imperfections limits the Q-factor to a few 10,000 \cite{Martinez2009_InP_RIE_QfactorL7, Srinivasan2003_InP_Qfactor}, and only recently the milestone of 100,000 has been achieved \cite{Yu2015, Crosnier2016_Qfactor}. Furthermore, there is no experimental demonstration (to the best of our knowledge) on the effect of p-doping on the Q-factor for PhC cavities.

\begin{figure*}
\centering\includegraphics[width=0.85\textwidth]{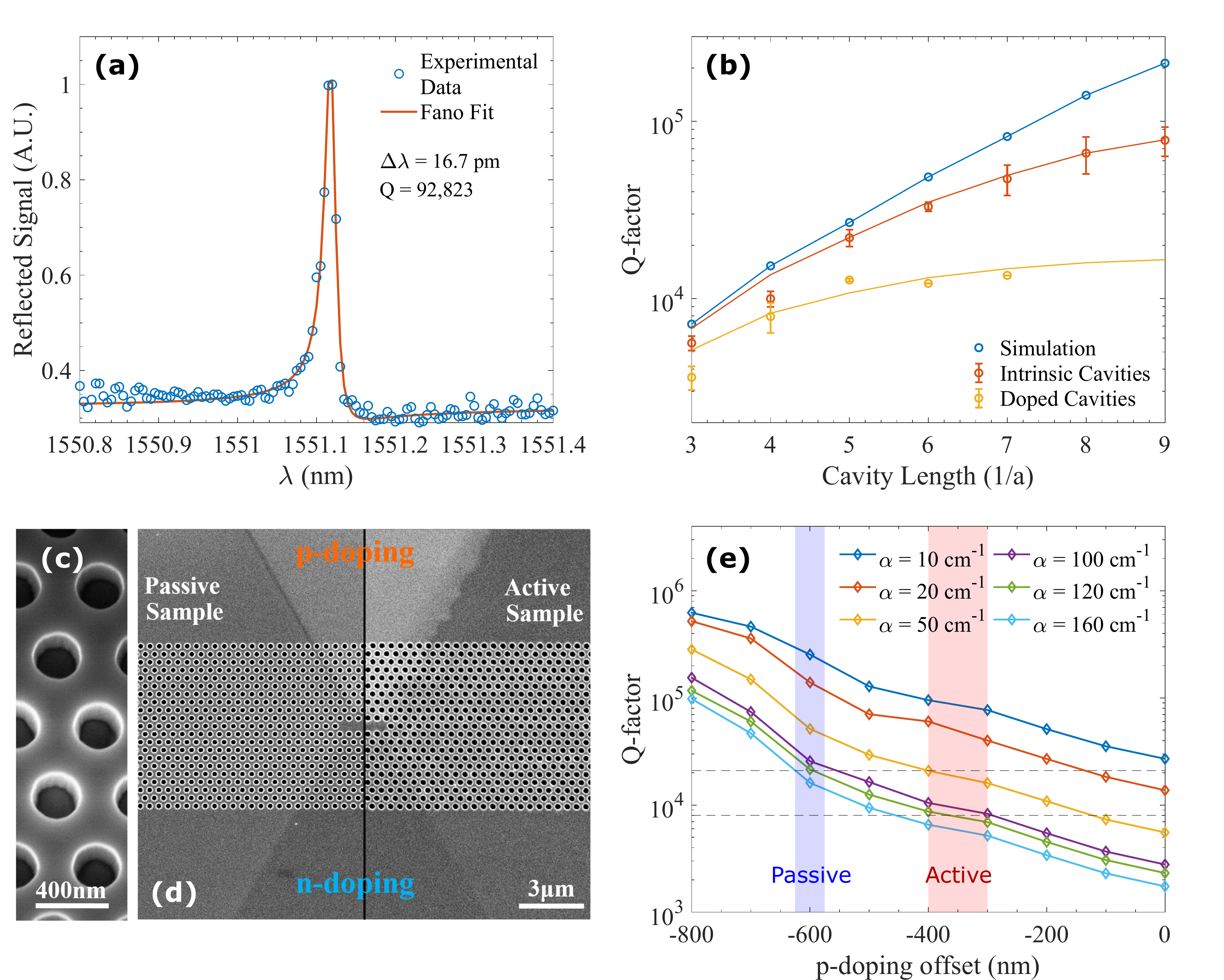}
\caption{(a) Resonant scattering spectrum of an L9 cavity. (b) The simulated intrinsic Q-factor, the experimental intrinsic Q-factor (without doping), and the total Q-factor (with doping) of LD cavities of different lengths. (c) A tilted SEM image of the fabricated PhC holes. (d) Passive (left) and active (right) PhC cavities. (e) Simulated Q-factor for an optimized L7 cavity as a function of the p-doping offset from the center of the cavity. The blue (red) shaded region shows the expected p-doping offset for a passive (active) cavity. The dashed lines are a guide to the eye.} \label{fig:Q-factor}
\end{figure*}

The Q-factor of passive InP LD cavities was experimentally measured by cross-polarization resonant scattering spectroscopy \cite{Galli2009}, which enables the direct measurement of the intrinsic Q-factor  (see a detailed description of the experimental setup in Note S2, Supporting Information). \textbf{Figure \ref{fig:Q-factor}}a, depicts the resonant scattering spectrum of an L9 cavity. The measured signal exhibits a characteristic Fano resonance \cite{Fano1961} due to the interference of the reflected pump light and the discrete cavity mode. A Fano lineshape fit is used to extract the linewidth ($\Delta\lambda$) and resonant wavelength ($\lambda_0$) of the fundamental mode. The Q-factor is calculated as $Q= \lambda_0/\Delta \lambda$.

The L9 cavity depicted in Figure \ref{fig:Q-factor}a exhibits a Q-factor of 93,000. However, simulating the same structure via the 3D FDTD method gives a Q-factor of 213,000, which indicates that there is a significant contribution of disorder due to fabrication imperfections. To quantify the effect of disorder on the Q-factor, several L3-L9 intrinsic cavities were characterized. For a laser structure that utilizes a lateral p-i-n structure, the total Q-factor will be further reduced due to the free carrier absorption of the p-doping region. As a result, doped cavities were characterized to evaluate the absorption due to p-doping. The total Q-factor of a PhC cavity is given by: 

\begin{equation} \label{eq.Qtotal}
    \frac{1}{Q_{tot}} =   \frac{1}{Q_{int}}  +  \frac{1}{Q_{dis}} +  \frac{1}{Q_{abs}} 
\end{equation}

where $Q_{tot}$ is the total Q-factor, $Q_{int}$ is the numerically calculated intrinsic Q-factor, $Q_{dis}$  is the disorder Q-factor associated with the radiation losses mainly due to structural imperfections, and $Q_{abs}$ represents an additional loss channel due to p-doping absorption. Absorption due to n-type doping is typically considered insignificant compared to the one of p-type doping \cite{Matsuo2018_review}, and thus was neglected in this analysis.

 In Figure \ref{fig:Q-factor}b, the theoretical and the experimental Q-factor of undoped and doped cavities are shown. Firstly, we calculate $Q_{dis}$ via equation \ref{eq.Qtotal}, using the simulated and experimental data of the intrinsic InP LD cavities. The fit is shown with the red solid line, and $Q_{dis}$ is estimated at 120,000 and is dependent on the disorder and the surface roughness of the etched PhC holes.  
 
 Finally, using the aforementioned values and the experimental data for the doped InP cavities, a new fit is used to calculate $Q_{abs}$ represented with the yellow solid line. In this fit, $Q_{abs}$ is estimated as 21,000 and is dependent on the p-doping levels and the proximity of the doping profile. This result demonstrates that such lateral p-i-n structures are limited by the optical losses induced by the p-doping. The fits show a good agreement of the theory with the experimental data, although, the Q-factor for smaller cavity lengths is overestimated since a larger part of the optical mode interacts with PhC holes enhancing the disorder-induced losses. 

Figure \ref{fig:Q-factor}c depicts a tilted SEM image of the PhC holes highlighting the surface roughness of the etched holes. An SEM image of a passive and active cavity is shown in the left and right part of Figure \ref{fig:Q-factor}d respectively. The Zn- and the Si-dopants provide enough contrast to visualize the doping profiles. The n-doping profile closely follows the DUV-mask design, as does the p-doping profile of the passive InP sample. However, the p-doping profile of the active sample is extended and exhibits some random wavy patterns attributed to the lower quality of InP regrown after dry etching, affecting the diffusion of the p-dopants. The extended p-region has a lower density of dopants, which is evident from the contrast in the SEM image.

 In Figure \ref{fig:Q-factor}e, 3D FDTD simulations show the calculated $Q_{abs}$ for different p-doping offsets from the center of the cavity and different p-doping absorption coefficients. As the p-doping profile is getting closer to the cavity center, the Q-factor drops, resembling a tri-exponential decay which relates to the overlap of the mode profile and the p-doping region. 

The absorption coefficient of the p-doping region is calculated as 120 cm$^{-1}$ using the p-doping offset extracted from the SEM image of the passive InP sample shown Figure \ref{fig:Q-factor}d. Consecutively, we can extrapolate the  $Q_{abs}$ of an active PhC cavity using the SEM of the active sample and the p-doping absorption coefficient. The shaded red and blue region of Figure \ref{fig:Q-factor}e shows the expected range of the p-doping offset for the active and passive sample respectively.  
As a result, the $Q_{abs}$ of an active laser cavity is estimated as 8000. Using the conventional semiconductor laser notation \cite{Coldren1995DiodeLA}, this corresponds to an average internal loss $\langle a_i \rangle$ of 17.1 cm$^{-1}$ (details on the simulation model are provided in Note S3, Supporting Information).
 
Overall, the Q-factor of a laser cavity is limited by the absorption of the p-doping region. We should note that although the Q-factor increases exponentially with the offset of the p-doping profile, the injection efficiency of the laser would massively drop due to the low mobility of the holes, discussed in section \ref{section:opticalPumping}. According to our measurements on LD lasers, the total Q-factor of a cavity should exceed 4000 to achieve lasing.

\subsection{Thermal Characteristics}

\begin{figure*}
\centering\includegraphics[width=0.85\textwidth]{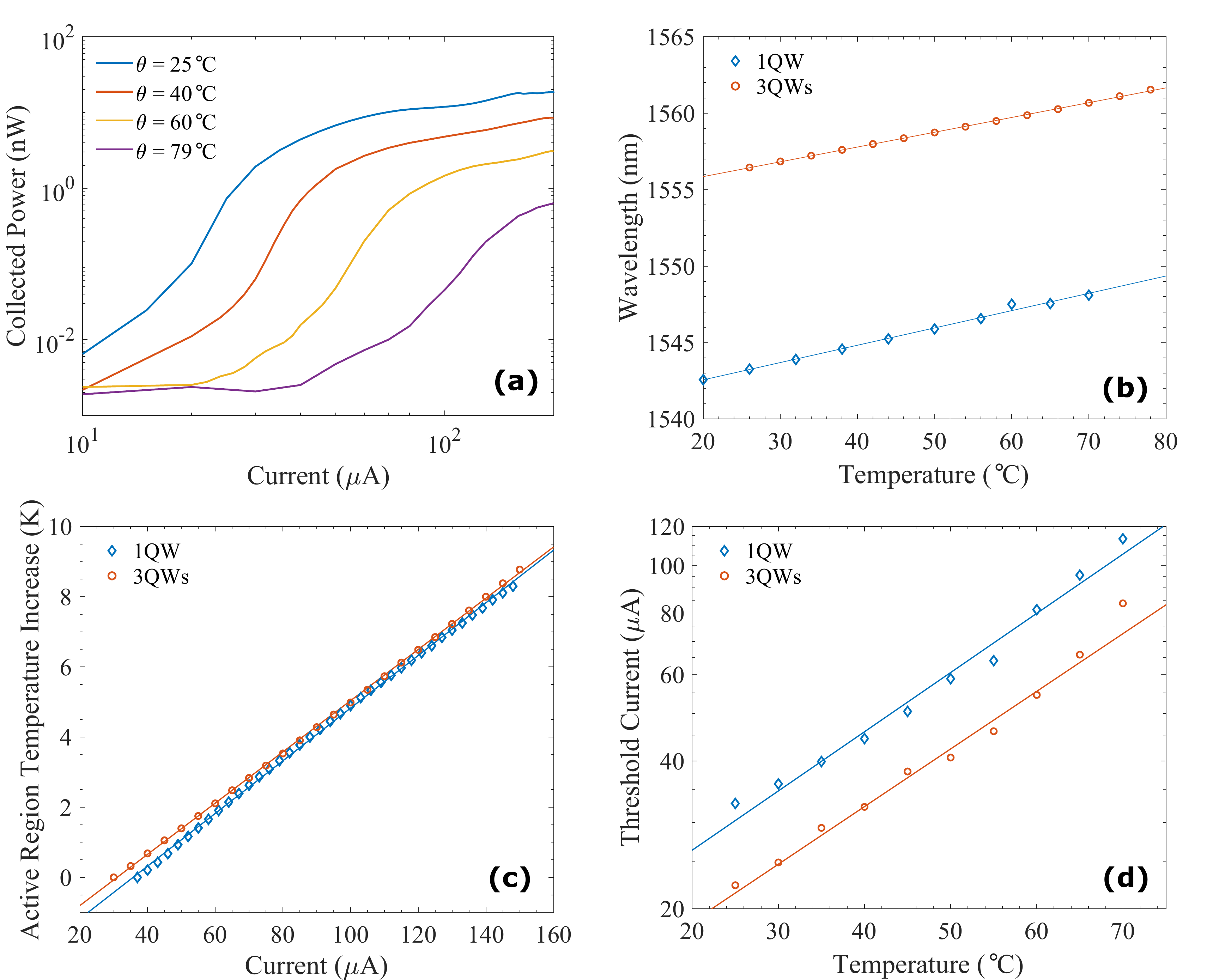}
\caption{Thermal characteristics of PhC lasers of one and three QWs. (a) L-I curve of an 3QW-based L7 laser for different heat sink temperatures. (b) The peak wavelength as a function of temperature. (c) Temperature increase of the active region for various pumping levels. (d) Threshold current dependence on heat sink temperature.} \label{fig:thermal}
\end{figure*}

Another important characteristic of lasers intended for inter- and intra-chip communication is their behavior at high temperatures and, in particular, the temperature dependence of the threshold current, which ultimately affects the power consumption, the output power, and the service life of the laser. In the following experiment, the thermal properties of lasers based on one and three QWs are investigated by adjusting the stage temperature. The temperature is varied from 20 \textdegree{}C to 79 \textdegree{}C via a thermoelectric cooler (TEC). The L-I curves of a standard 3QW-L7 laser for four different heat sink temperatures are shown in \textbf{Figure \ref{fig:thermal}}a. Lasing is achieved for up to 79 \textdegree{}C, which is the upper limit of the used TEC.

The laser wavelength is linearly dependent on the heat sink temperature redshifting by 0.1 nm/K for both one and three QWs which is shown in Figure \ref{fig:thermal}b. This linear increase is due to the increased refractive index \cite{Geng2018_thermal_effect_in_refractive_index}. Subsequently, one can calculate the temperature increase in the active region of the laser under varying pumping levels. The active temperature increase is depicted in Figure \ref{fig:thermal}c and was approximately 8.5\textdegree{}C at 150 \textmu A and had an average slope of 84 K/mA for both lasers. The relatively high thermal conductivity is one of the advantages of the BH technology since the poorly thermally-conductive InGaAsP active region (68 W m$^{-1}$K$^{-1}$) is embedded in the better thermally-conductive InP membrane  (4.2 W m$^{-1}$K$^{-1}$) \cite{Matsuo2010_optical_BH}. The generated heat under varying pumping is not dependent on the dissipated power due to ohmic heating, but due to the absorption of the increased optical power density circulating the cavity. This was confirmed by the comparison of the lasing wavelength under optical pumping, discussed in section \ref{section:opticalPumping}. Moreover, less powerful lasers operate at lower effective temperatures. One example would be the L3 laser showed in Figure \ref{fig.L3} that exhibited a much smaller active temperature increase with a slope of 43 K/mA.

As the temperature of the heat sink rise, the injection efficiency and the gain drops \cite{Coldren1995DiodeLA, Piprek2000_thermal_properties_Gain_Auger_Leakage}, while the Auger recombination-induced losses increase \cite{Fuchs1993_Auger_QWs}. This leads to an increase in the laser threshold current. The sensitivity of the threshold current concerning temperature can be quantified using a characteristic temperature $T_0$ via the commonly used empirical relation \cite{Pankove1968}:

\begin{equation}
    I_{th}(T)=I_0 exp\left( \frac{T}{T_0} \right) \label{eq:exponential_threshold}
\end{equation}

The evolution of laser threshold current to heat sink temperature is shown in figure \ref{fig:thermal}d for the 1QW- and 3QW-based lasers. A good fit with equation \ref{eq:exponential_threshold} was found giving a characteristic temperature $T_0$ of 35 \textdegree{}C for both 1QW and 3QWs lasers. This value is in the low end for InGaAsP-based QW lasers \cite{Coldren1995DiodeLA,OGorman1992}, attributed to the poor heat dissipation of the free-floating PhC membrane.

The thermal properties of these devices can be greatly improved if the PhC slab is surrounded by a low-index material like SiO$_2$ or polymer acting as a heatsink \cite{Bazin2014_ranieriThermal}.


\subsection{Injection Efficiency - Comparison between Optical and Electrical pumping} \label{section:opticalPumping}

One of the main effects limiting the efficiency of the laterally doped 2D PhC nanolasers is the low injection efficiency estimated to be in the order of 1-10\% from data of previously demonstrated lasers \cite{Takeda2021, Matsuo2013}. The lateral doping geometry offers the possibility for both electrical and optical pumping, and thus a comparison between the two pumping schemes was performed to understand the limiting factors on the efficiency. 

\begin{figure*}
\centering\includegraphics[width=1\textwidth]{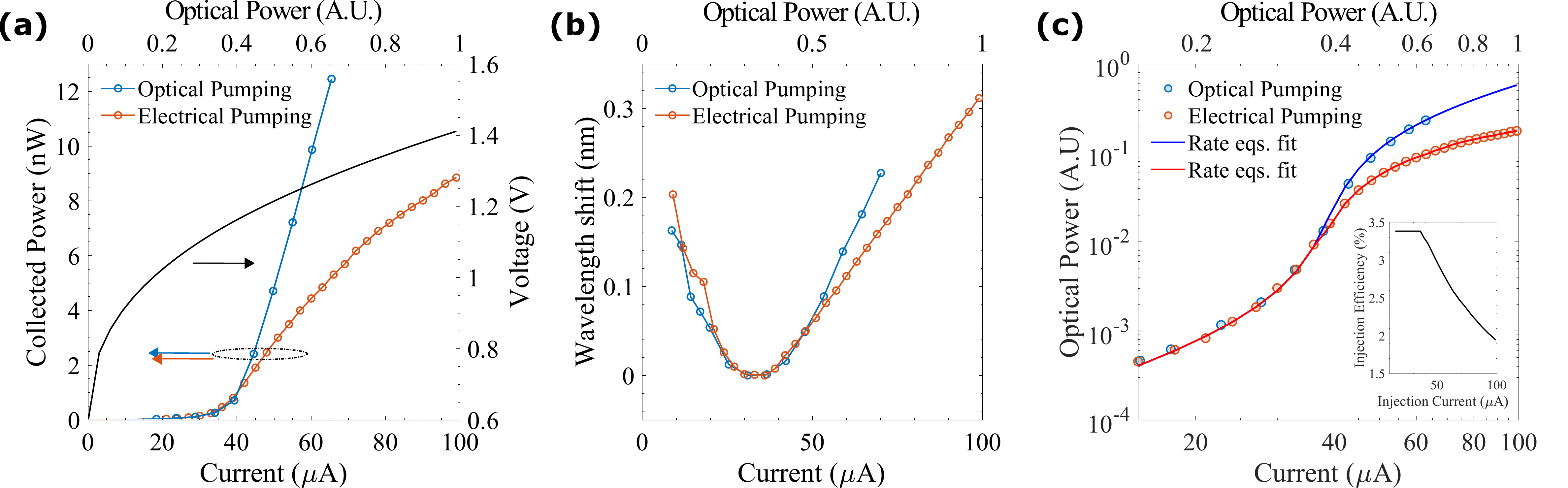}
\caption{(a) Comparison of Input-Output curves of an L5 laser under optical and electrical pumping. (b) Comparison of the wavelength evolution. (c) Fitting of the conventional and the modified rate equations on laser I-O curves for optical and electrical pumping respectively. The inset shows the modified injection efficiency vs the injection current for the electrical pumping scheme .} \label{fig:optPumping}
\end{figure*}

In \textbf{Figure \ref{fig:optPumping}}(a), the L-I-V curves and the L-L curve for the electrical and optical pumping of a 1QW-L5 laser are shown. In the optical pumping scheme, a 1310 nm pump laser was coupled to a single-mode fiber, while the same 50x objective was used for pumping and collecting. The optical pump power was normalized to match the laser threshold in both pumping schemes. We observe, however, that the output power is much lower after the threshold for the case of electrical injection. This effect is attributed to a drop in the injection efficiency as the applied voltage and current increase, and it is not related to heating since the spectral evolution of the laser peak is very similar for both pumping schemes. The wavelength evolution is shown in Figure \ref{fig:optPumping}b, demonstrating that the heating after the threshold is mainly attributed to the optical field circulating the cavity and not the ohmic heating. 

pump was coupled to a single-mode fiber, while the same 50x objective was used for pumping and collecting. The signal was separated from the pump using a wavelength division multiplexer and was sent to an OSA to record the optical spectrum and the output power. The optical pump power was normalized to match the laser threshold of the electrical pumping scheme, however, the output power is much lower after the threshold for the electrical injection. This effect is attributed to the drop of the injection efficiency as the voltage and current increase, and it is not related to heating, since the spectral evolution of the laser peak is very similar for both pumping schemes.   

To quantify this effect, we analyze the experimental results using the conventional laser rate equations \cite{Coldren1995DiodeLA}:

\begin{subequations}
				\begin{align}
    			    \frac{dN}{dt} =& R_{pump} - \left[ \frac{1}{\tau_r} + \frac{1}{\tau_{nr}}\right] N - u_g g(N) N_p \label{laser_rate_eq1}\\
    			    \frac{dN_p}{dt} =& \left[ \Gamma u_g g(N) - \frac{1}{\tau_p} \right] N_p + \Gamma \beta \frac{N}{\tau_r} \label{laser_rate_eq2}
    			    \end{align}\label{laser_rate_eq}
\end{subequations}

where $N$ and $N_p$ is the carrier density and photon density inside the cavity, $R_{pump}$ is the carrier pumping rate, $\tau_{r}$, $\tau_{nr}$ and $\tau_{p}$, is the radiative, non-radiative and photon lifetime respectively. $\Gamma$ is the confinement factor, $u_g$ is the group velocity, and $g(N)$ is the gain of the active material. The pumping rate depends on the pumping method and was defined as: 

\begin{equation}
  R_{pump}=\begin{cases}
    {\eta_e I}/{q V} ,& \text{for Electrical Pumping.}\\
    \eta_o P_{in}/ \hbar \omega_p V ,& \text{for Optical Pumping}.
  \end{cases}
\end{equation}
where $\eta_e$ is the electrical injection efficiency, $I$ is the injected current, $q$ is the elementary charge and $V$ is the volume of the active material. For the optical pumping rate, $\eta_o$ is the optical pumping efficiency, $P_{in}$ is the optical incident pump power, and $ \hbar \omega_p$ is the energy per pump photon.

    In Figure \ref{fig:optPumping}c the experimental data and the rate equation fits for optical and electrical pumping are plotted in a logarithmic scale. For both fits, the same parameters of the laser are used. Namely, $\tau_{r}$ is 2 ns and $\tau_{nr}$ is 10 ns respectively, the group velocity $u_g = c/n_g$ is calculated based on a group index $n_g$ of 3.5. A logarithmic gain model given by $g(N) = g_0 ln(N/N_0)$ was used to describe the gain of the QW where the gain coefficient is $g_0$ is 2000 cm$^{-1}$, and the transparency carrier density $N_0$ is $8.7\cdot10^{17}$ cm$^{-3}$. The confinement factor $\Gamma$ is 4\%. The photon lifetime can be calculated as $\tau_{p} = Q/\omega$ where $\omega$ is the laser frequency, and Q is the total Q-factor of the laser cavity. Following the discussion of subsection \ref{ch.Q-factor}, the Q-factor is chosen as 8000. 

A good agreement of the conventional rate equations fitting and the experimental data was found for the optical pumping scheme, where the $\beta$-factor was calculated as 0.03 and is on par with previously reported results \cite{ Matsuo2013, Xue2016}. 
For the case of electrical pumping, however, the conventional rate equations could not accurately describe the behavior after threshold in any parameter combination which is attributed to a voltage-dependent drop of the injection efficiency. As a result, the rate equations were modified by modeling $\eta_e$ as a function of the injected current as shown in the inset of Figure \ref{fig:optPumping}c. From this fit, the injection efficiency was calculated as 0.03, decreasing down to 60\% of its original value. Similar injection efficiency drop was observed in all lasers and typically happens around 1.1 V and 1.3 V, which are the turn-on voltages of the QW barrier and InP layer, respectively.

\begin{figure*}
\centering\includegraphics[width=0.85 \textwidth]{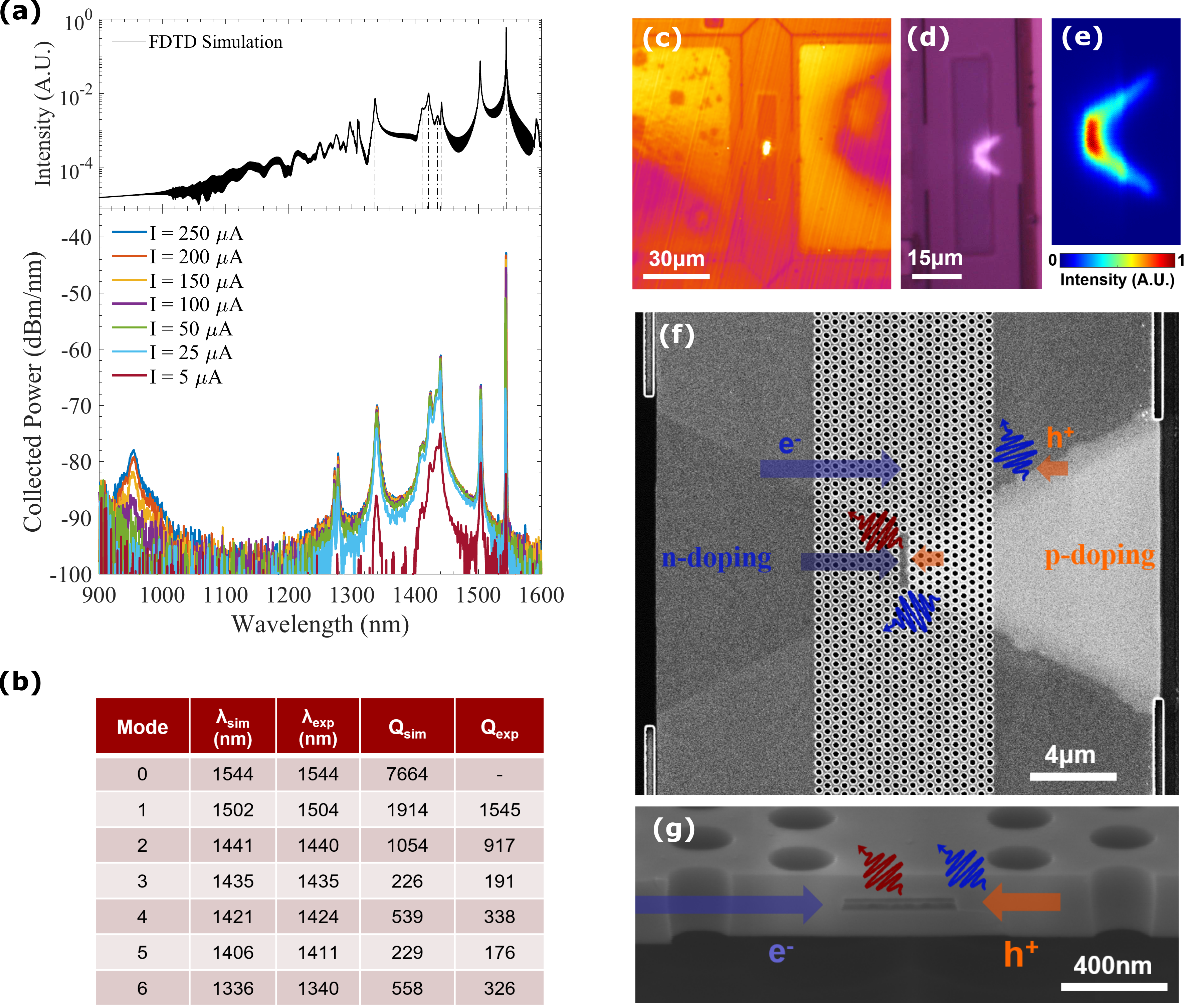}
\caption{(a) The measured optical spectrum of an L5 laser for different injection currents (bottom) and the simulated spectrum (top). The dash-dotted lines are a guide for the eye. (b) Table comparing the experimental and the simulation results. (c) A microscope image of a PhC laser taken by an InGaAs camera. (d) A microscope image using a Si camera. (e) Heatmap of the emission profile of the InP closely resembling the lateral p-doping profile. (f) Artificially colored SEM image of a laser depicting the flow of electrons and holes and the photon emission from the QWs and the p-i interface. (g) Artificially colored SEM image of the cross-sectional view of the BH active material confined in a W1 waveguide.} \label{fig:Leakage}
\end{figure*}

To further understand this effect, the optical spectra of the laser for different pumping conditions are shown in \textbf{Figure \ref{fig:Leakage}}a revealing the higher-order modes. To increase the collection efficiency the vertically scattered light was coupled to a multi-mode fiber whose alignment was based on maximizing the collection of the lasing peak at 1544 nm. A slight adjustment in the fiber alignment can affect the relative intensity of the modes, although their intensity is mainly dependent on the Q-factor, and the far-field mode overlap with the objective. The laser exhibited a threshold current of 35 \textmu A, however, significant emission from higher-order cavity modes can be observed even at 5 \textmu A. Above threshold, the emission of the higher-order modes is mostly clamped, however, as the applied voltage and the injection current are increased a peak at 950 nm is observed that is attributed to spontaneous emission from InP.

The 3D-FDTD simulated spectrum of the structure is shown in the upper part of Figure \ref{fig:Leakage}a. There is a good agreement of the numerical and experimental values for the resonant peaks and the Q-factor up to the sixth-excited cavity mode, summarized in the table of Figure \ref{fig:Leakage}b. The spontaneously emitted light in the higher-order modes closely resembles the theoretical cold cavity, while the fundamental mode assumed the cold cavity Q-factor at $0.4\cdot I_{th}$ that can be considered the current required to reach transparency. 

The InP emission can be spatially resolved via the microscope setup. In Figure \ref{fig:Leakage}c, a microscope image of a running laser is captured by an InGaAs camera. Using a Si camera, photon emission from the interface of the p-doping region was observed as shown in Figure \ref{fig:Leakage}d, demonstrating that there is a significant leakage current that leads to a low injection efficiency. Figure \ref{fig:Leakage}e depicts a heatmap of the InP electroluminescence that clearly outlines the p-doping interface. The carrier recombination occurs near the p-i interface due to the low mobility of the holes. The mechanism of the leakage is depicted in Figure \ref{fig:Leakage}(f-g) where the top and cross-sectional view of an electron micrograph is artificially colored. In these pictures, the electrons and the holes are represented with blue and orange arrows, while the QW and InP photons are represented with the red and blue wavy arrows, respectively. Leakage paths have been identified both in the vertical and lateral direction and thus the shape and offset of the p-doping profile have to be optimized. As discussed in section \ref{ch.Q-factor}, there is a trade-off between the Q-factor and the injection efficiency, although an optimal value has yet to be extracted. The limit of the p-doping offset was experimentally determined as 800 nm where the injection efficiency is too low to achieve lasing.

\section{Conclusion}

In this work, we reported continuous-wave electrical operation of an L3 photonic crystal nanolaser with an ultra-low threshold current of 10.2 \textmu A at room temperature, emitting at 1540 nm. The active material consists of a quantum well in a buried heterostructure region where the carriers are injected via a lateral p-i-n junction. Detailed information on the design method and the nanofabrication process was given. The effect of the disorder and the p-doping on Q-factor was experimentally quantified via cross-polarization resonant scattering measurements on InP line defect cavities. The absorption coefficient of the p-doping region was deduced to be 120 cm\textsubscript{-1}  by combining the experimental results with FDTD simulations. Thus, the Q-factor of a laser cavity is estimated at 8000, dominated by the losses due to the p-doping absorption. Furthermore, the thermal behavior of one and three quantum well lasers was characterized. The temperature increase of the active region was found to be under 10K under normal operating conditions and is mainly dependent on the intensity of the circulating optical field and not ohmic losses. However, the threshold was shown to be strongly affected by the heat sink temperature with a characteristic temperature of 35 degrees which is related to a reduction of the injection efficiency. The injection efficiency drop was further investigated via a comparison between the optical and electrical pumping scheme and was quantified using the laser rate equations. Finally, InP emission from the p-i interface was spectrally and spatially resolved proving that there is a significant leakage current that limits the injection efficiency of laterally doped photonic crystal nanolasers.

\medskip
\textbf{Acknowledgements} \par 
Authors gratefully acknowledge funding by Villum Fonden via the NATEC Center of Excellence (Grant No. 8692), the European Research Council (ERC) under the European Union’s Horizon 2020 Research and Innovation Programme (Grant No. 834410 FANO), and the Danish National Research Foundation (Grant No. DNRF147 NanoPhoton).

\medskip
\textbf{Disclosures}\par
The authors declare no conflicts of interest.

\medskip
\textbf{Data Availability} \par
 The data that support the findings of this study are available from the corresponding author upon reasonable request.

\bibliography{bibtex.bib}

\end{document}